\documentclass[12pt]{article}
\usepackage{epsfig}

\newcommand\pubnumber{TTP01-04}
\newcommand\pubdate{January 15, 2001}
\newcommand\hepnumber{hep-ph/0101147}
\def\lapprox{\lower .7ex\hbox{$\;\stackrel{\textstyle <}{\sim}\;$}}
\def\gapprox{\lower .7ex\hbox{$\;\stackrel{\textstyle >}{\sim}\;$}}

\def\e{\epsilon}
\def\d{{\rm d}}

\def\sab{s_{12}}
\def\sac{s_{13}}
\def\sbc{s_{23}}

\def\sabc{s_{123}}

\def\csumb{$^a$ Theory Division, CERN, CH-1211 Geneva 23, Switzerland\\
$^b$ Institut f\"ur Theoretische Teilchenphysik,
Universit\"at Karlsruhe,\\ D-76128 Karlsruhe, Germany\\
$^c$ Dipartimento di Fisica,
    Universit\`{a} di Bologna and INFN, Sezione di 
    Bologna,  I-40126 Bologna, Italy} 

\def\Title#1{\begin{center} {\Large\bf #1 } \end{center}}
\def\Author#1{\begin{center}{ \sc #1} \end{center}}
\def\Address#1{\begin{center}{ \it #1} \end{center}}

\newcommand\pubblock{\rightline{\begin{tabular}{l} \pubnumber\\
         \pubdate\\ \hepnumber \end{tabular}}}
\newenvironment{Abstract}{\begin{quotation}  }{\end{quotation}}
\newenvironment{Presented}{\begin{quotation} \begin{center} 
             Presented at the\end{center}
      \begin{center}\begin{large}}{\end{large}\end{center} \end{quotation}}
\def\Acknowledgments{\bigskip  \bigskip \begin{center}
          \large\bf Acknowledgments\end{center}}

\makeatletter
\def\section{\@startsection{section}{0}{\z@}{5.5ex plus .5ex minus
 1.5ex}{2.3ex plus .2ex}{\large\bf}}
\def\subsection{\@startsection{subsection}{1}{\z@}{3.5ex plus .5ex minus
 1.5ex}{1.3ex plus .2ex}{\normalsize\bf}}
\def\subsubsection{\@startsection{subsubsection}{2}{\z@}{-3.5ex plus
-1ex minus  -.2ex}{2.3ex plus .2ex}{\normalsize\sl}}

\renewcommand{\@makecaption}[2]{%
   \vskip 10pt
   \setbox\@tempboxa\hbox{\small #1: #2}
   \ifdim \wd\@tempboxa >\hsize     
       \small #1: #2\par          
     \else                        
       \hbox to\hsize{\hfil\box\@tempboxa\hfil}
   \fi}

 \def\citenum#1{{\def\@cite##1##2{##1}\cite{#1}}}
 
\newcount\@tempcntc
\def\@citex[#1]#2{\if@filesw\immediate\write\@auxout{\string\citation{#2}}\fi
  \@tempcnta\z@\@tempcntb\m@ne\def\@citea{}\@cite{\@for\@citeb:=#2\do
    {\@ifundefined
       {b@\@citeb}{\@citeo\@tempcntb\m@ne\@citea\def\@citea{,}{\bf ?}\@warning
       {Citation `\@citeb' on page \thepage \space undefined}}%
    {\setbox\z@\hbox{\global\@tempcntc0\csname b@\@citeb\endcsname\relax}%
     \ifnum\@tempcntc=\z@ \@citeo\@tempcntb\m@ne
       \@citea\def\@citea{,}\hbox{\csname b@\@citeb\endcsname}%
     \else
      \advance\@tempcntb\@ne
      \ifnum\@tempcntb=\@tempcntc
      \else\advance\@tempcntb\m@ne\@citeo
      \@tempcnta\@tempcntc\@tempcntb\@tempcntc\fi\fi}}\@citeo}{#1}}
\def\@citeo{\ifnum\@tempcnta>\@tempcntb\else\@citea\def\@citea{,}%
  \ifnum\@tempcnta=\@tempcntb\the\@tempcnta\else
  {\advance\@tempcnta\@ne\ifnum\@tempcnta=\@tempcntb \else\def\@citea{--}\fi
    \advance\@tempcnta\m@ne\the\@tempcnta\@citea\the\@tempcntb}\fi\fi}
\makeatother

\def\beq{\begin{equation}}
\def\eeq#1{\label{#1}\end{equation}}
\def\eeqn{\end{equation}}

\newenvironment{Eqnarray}%
   {\arraycolsep 0.14em\begin{eqnarray}}{\end{eqnarray}}
\def\beqa{\begin{Eqnarray}}
\def\eeqa#1{\label{#1}\end{Eqnarray}}
\def\eeqan{\end{Eqnarray}}

\let\bar=\overbar

\def\Dslash{\not{\hbox{\kern-4pt $D$}}}
\def\dslash{\not{\hbox{\kern-2pt $\del$}}}

\def\msb{{\bar{\ssstyle M \kern -1pt S}}}

\def\lsim{\mathrel{\raise.3ex\hbox{$<$\kern-.75em\lower1ex\hbox{$\sim$}}}}
\def\gsim{\mathrel{\raise.3ex\hbox{$>$\kern-.75em\lower1ex\hbox{$\sim$}}}}

\begin{document}
\begin{titlepage}
\pubblock

\vfill
\def\thefootnote{\fnsymbol{footnote}}
\Title{Progress on two-loop non-propagator integrals}
\vfill
\Author{\underline{T.~Gehrmann}$^{a,b}$ and E.~Remiddi$^c$}
\Address{\csumb}
\vfill
\begin{Abstract}
At variance with fully inclusive quantities, which have been computed 
already at the two- or three-loop level, most exclusive observables are 
still known only at one loop, as further progress was hampered 
up to very recently
 by the greater computational problems encountered in the study 
of multi-leg amplitudes beyond one loop. We discuss the progress made lately 
in the evaluation of two-loop multi-leg integrals, with particular emphasis
on two-loop four-point functions. 
\end{Abstract}
\vfill
\begin{Presented}
5th International Symposium on Radiative Corrections \\ 
(RADCOR--2000) \\[4pt]
Carmel CA, USA, 11--15 September, 2000
\end{Presented}
\vfill
\end{titlepage}
\def\thefootnote{\arabic{footnote}}
\setcounter{footnote}{0}

\section{Introduction}

Precision applications of particle physics phenomenology often demand 
theoretical predictions at the next-to-next-to-leading order 
in perturbation theory. Corrections at this order are known
for many inclusive observables, such as total cross sections or sum rules,
which correspond from a technical point of view to propagator-type 
Feynman amplitudes.
For  $2\to 2$ scattering and $1\to 3$ decay 
processes, the calculation of next-to-next-to-leading order corrections 
is a yet outstanding task. 
One of the major ingredients for these 
calculations are the two-loop virtual corrections to the 
corresponding four-point Feynman amplitudes. 
Depending on the process under consideration, these calculations 
require two-loop four-point functions with massless internal 
propagators and all legs on-shell (high energy limit of 
Bhabha scattering, hadronic two-jet production) or one leg off-shell
(three-jet production and event shapes in electron--positron annihilation, 
two-plus-one-jet production in deep inelastic scattering, hadronic 
vector-boson-plus-jet production). 

During the past two years, many new results on two-loop four-point
functions
became available, thus enabling the first calculations of two-loop 
virtual corrections to  $2\to 2$ scattering processes. A variety of 
newly developed techniques made this progress possible. In this talk, we
describe these new techniques and their applications, and we summarise
recent results.
In an outlook, we discuss the remaining steps to be taken
towards the completion of next-to-next-to-leading order calculations 
of $2\to 2$ scattering and $1\to 3$ decay 
processes.

\section{New technical developments}

Using dimensional regularization~\cite{dreg,hv} 
with $d=4-2\e$ dimensions as regulator for
ultraviolet and infrared divergences, the integrals appearing 
in the calculation of two-loop corrections take the generic form
\begin{equation}
I(p_1,\ldots,p_n) = 
 \int \frac{\d^d k}{(2\pi)^d}\frac{\d^d l}{(2\pi)^d} 
\frac{1}{D_1^{m_1}\ldots D_{t}^{m_t}} S_1^{n_1} 
\ldots S_q^{n_q} \; ,
\label{eq:generic}
\end{equation}
where the $D_i$ are massless scalar propagators, depending on $k$, $l$ and the 
external momenta $p_1,\ldots,p_n$ while $S_i$ are scalar products
of a loop momentum with an external momentum or of the two loop
momenta.  The topology (interconnection of
propagators and external momenta) of the integral is uniquely 
determined by specifying the set $(D_1,\ldots,D_t)$
of $t$ different propagators in the graph. The integral itself is then
specified by the powers $m_i$ of all propagators and by the selection 
$(S_1,\ldots,S_q)$  of scalar products and their powers $(n_1,\ldots,n_q)$ 
(all the $m_i$ are positive integers greater or equal to 1, while the 
$n_i$ are greater or equal to 0). 
Integrals of the same topology with the same dimension $r=\sum_i m_i$ 
of the denominator and same total number $s=\sum_i n_i$ of scalar products 
are denoted as a class of integrals $I_{t,r,s}$. The integration measure and 
scalar products appearing the above expression are in Minkowskian space, 
with the usual causal prescription for all propagators. The loop
integrations are carried out for arbitrary space-time dimension $d$, 
which acts as a regulator for divergences appearing due to the
ultraviolet or infrared behaviour of the integrand. 
For each topology appearing in the 
calculation, a sizable number of different scalar integrals has to be 
computed. 

Recent progress in the computation of two-loop  corrections 
to four-point amplitudes 
 was based on three technical developments: 
an efficient procedure to reduce the large number of 
different scalar integrals 
to a very limited number of so-called master integrals, new techniques for 
the computation of these master integrals, and a new class of functions
(harmonic polylogarithms), which can be extended to suit the needs of 
a particular calculation. We discuss these developments in the following.

\subsection{Reduction to master integrals}

The number $N(I_{t,r,s})$ of the integrals grows quickly as $r, s$ 
increase, but the integrals are related among each other
by various identities. 
One class of identities follows from the fact that the integral over the 
total derivative with respect to any loop momentum vanishes in
dimensional regularization
\begin{equation}
\int \frac{\d^d k}{(2\pi)^d} \frac{\partial}{\partial k^{\mu}}
J(k,\ldots)  = 0,
\end{equation} 
where $J$ is any combination of propagators, scalar products
and loop momentum vectors. $J$ can be a vector or tensor of any rank. 
The resulting identities~\cite{hv,chet} 
are called integration-by-parts (IBP)
identities.

In addition to the IBP identities, one can also exploit the fact that
all integrals under consideration are Lorentz scalars (or, perhaps 
more precisely, ``$d$-rotational'' scalars) , which are
invariant under a Lorentz (or $d$-rotational) transformation of the 
external momenta~\cite{gr}. These Lorentz invariance (LI) identities 
are obtained from:
\begin{equation}
\left(p_1^{\nu}\frac{\partial}{\partial
    p_{1\mu}} - p_1^{\mu}\frac{\partial}{\partial
    p_{1\nu}} + \ldots
 + p_n^{\nu}\frac{\partial}{\partial
    p_{n\mu}} - p_n^{\mu}\frac{\partial}{\partial
    p_{n\nu}}\right) I(p_1,\ldots,p_n) = 0 \;. 
\label{eq:li}
\end{equation}

In the case of two-loop four-point functions, one has a total of 13
equations (10 IBP + 3 LI) for each integrand corresponding to an 
integral of class $I_{t,r,s}$, relating integrals of the same topology
with up to $s+1$ scalar products and $r+1$ denominators, plus integrals
of simpler topologies ({\it i.e.}~with a smaller number of different 
denominators). 
The 13 identities obtained starting from an integral $I_{t,r,s}$ do contain
integrals of the following types:
\begin{itemize}
\item $I_{t,r,s}$: the integral itself. 
\item $I_{t-1,r,s}$: simpler topology. 
\item $I_{t,r+1,s}, I_{t,r+1,s+1}$ : same topology, more complicated than
  $I_{t,r,s}$.  
\item $I_{t,r-1,s}, I_{t,r-1,s-1}$: same topology, simpler than 
   $I_{t,r,s}$.  
\end{itemize}
Quite in general, single identities of the above kind can be used 
to obtain the reduction of $I_{t,r+1,s+1}$ or $I_{t,r+1,s}$ integrals 
in terms of $I_{t,r,s}$ and simpler integrals - rather than to 
get information on the $I_{t,r,s}$ themselves. 

If one considers the set of all the identities obtained starting from 
the integrand of all the $N(I_{t,r,s})$ integrals of class $I_{t,r,s}$, 
one obtains 
$(N_{{\rm     IBP}}+ N_{{\rm LI}}) N(I_{t,r,s})$ identities 
which contain $N(I_{t,r+1,s+1})+N(I_{t,r+1,s})$ 
integrals of more complicated structure. It was first noticed by S.\
Laporta~\cite{laporta}
that with increasing $r$ and $s$
the number of identities grows faster than the number
of new unknown integrals. 
As a consequence, if for a given $t$-topology one considers the set of 
all the possible equations obtained by considering all the integrands up to 
certain values $r^*, s^*$ of $r, s$, for large enough $r^*, s^*$ 
the resulting system of equations, apparently overconstrained,
can be used for 
expressing the more complicated integrals, with greater values of $r, s$ 
in terms of simpler ones, with smaller values of $r, s$. 
An automatic procedure to perform
this reduction by means of computer algebra using FORM~\cite{form} and 
MAPLE~\cite{maple}
is discussed in more detail in~\cite{gr}.

For any given four-point two-loop topology, 
this procedure can result either in a reduction
 towards a small number (typically one or two) of integrals of the
topology under consideration and integrals of simpler 
topology (less different denominators), or even in a complete 
reduction of all integrals of the topology under consideration
towards integrals with simpler topology.
Left-over integrals of the topology under consideration are called 
irreducible master integrals or just 
master integrals. 

\subsection{Computation of master integrals}

The IBP and LI identities allow to
express integrals of the form (\ref{eq:generic}) as a linear
combination of a few master integrals,  i.e.\ integrals which are 
not further reducible, but have 
to be computed by some different method. 

For the case of massless two-loop 
four-point functions, several techniques have been proposed in the
literature, such as for example the application of a Mellin--Barnes
transformation to all propagators~\cite{onshell1} 
or the negative dimension
approach~\cite{glover}. Both techniques rely on an explicit integration 
over the loop momenta, with differences mainly in the representation
used for the propagators.

A method for the analytic
computation of master integrals avoiding the explicit
integration over the loop momenta is to derive differential equations in 
internal propagator masses or in external momenta for the master integral, 
and to solve these with appropriate boundary conditions. 
This method has first been suggested by Kotikov~\cite{kotikov} to relate 
loop integrals with internal masses to massless loop integrals. 

It has been elaborated in detail and generalized to differential 
equations in external momenta in~\cite{remiddi}; first 
applications were presented in~\cite{appl}.
In the case of four-point functions with one external off-shell leg
and no internal masses, one has three independent
invariants, resulting in three differential equations.

The derivatives in the invariants $s_{ij}=(p_i+p_j)^2$ 
can be expressed by derivatives in the external momenta:
\begin{eqnarray}
\sab \frac{\partial}{\partial \sab} & = & \frac{1}{2} \left( +
p_1^{\mu} \frac{\partial}{\partial p_1^{\mu}} +
p_2^{\mu} \frac{\partial}{\partial p_2^{\mu}} -
p_3^{\mu} \frac{\partial}{\partial p_3^{\mu}}\right)\;, \nonumber \\
\sac \frac{\partial}{\partial \sac} & = & \frac{1}{2} \left( +
p_1^{\mu} \frac{\partial}{\partial p_1^{\mu}} -
p_2^{\mu} \frac{\partial}{\partial p_2^{\mu}} +
p_3^{\mu} \frac{\partial}{\partial p_3^{\mu}}\right)\;,\nonumber \\
\sbc \frac{\partial}{\partial \sbc} & = & \frac{1}{2} \left( - 
p_1^{\mu} \frac{\partial}{\partial p_1^{\mu}} +
p_2^{\mu} \frac{\partial}{\partial p_2^{\mu}} +
p_3^{\mu} \frac{\partial}{\partial p_3^{\mu}}\right)\;. 
\label{eq:derivatives} 
\end{eqnarray}

It is evident that acting with the right hand  sides of (\ref{eq:derivatives}) 
on a master integral $I_{t,t,0}$
will, after interchange of 
derivative and integration, yield 
a combination of 
integrals of the same type as appearing in the IBP and LI identities for 
$I_{t,t,0}$, including integrals of type $I_{t,t+1,1}$ and 
$I_{t,t+1,0}$. Consequently, the scalar derivatives (on left hand side of 
(\ref{eq:derivatives}))
of  $I_{t,t,0}$ can be expressed by a linear combination of 
integrals up to  $I_{t,t+1,1}$ and 
$I_{t,t+1,0}$.
These can all be reduced (for topologies containing only one 
master integral) to $I_{t,t,0}$ 
and to integrals of simpler topology
by applying the IBP and LI identities. As a result, we obtain 
for the master integral $I_{t,t,0}$ 
an inhomogeneous linear
first order differential equation in each invariant. For topologies with 
more than one master integral, one finds a coupled system of 
first order differential equations.
The inhomogeneous term in these differential equations contains only 
topologies simpler than $I_{t,t,0}$, which are considered to be known 
if working in a bottom-up approach. 

The master integral $I_{t,t,0}$ is obtained by matching the general 
solution of its differential equation to an appropriate boundary
condition. Quite in general, finding a boundary condition is 
a  simpler problem than evaluating the whole
integral, since it depends on a smaller number of kinematical
variables. In some cases, the boundary condition can even be
determined from the differential equation itself. 

To solve the differential equations for two-loop four-point functions 
with one off-shell leg~\cite{gr,gr1}, we 
express
the system of differential equations for 
any master integral in the variables 
$\sabc = \sab + \sac + \sbc$, $y=\sac/\sabc$ and $z=\sbc/\sabc$.
We
obtain a homogeneous equation in $\sabc$, and inhomogeneous equations 
in $y$ and $z$. Since $\sabc$ is the only quantity carrying a mass 
dimension, 
the corresponding differential equation is nothing but the rescaling 
relation obtained by investigating the behaviour of the master integral 
under a rescaling of all external momenta by a constant factor. 
The master integral can be determined by solving one of the
inhomogeneous equations, the second equation can then serve as a check 
on the result.

In the
$y$  differential equation
for the master integral under consideration, the
coefficient of the homogeneous term as well as the full inhomogeneous
term (coefficients and subtopologies) are then
expanded as a series in $\e$. From the leading coefficient of the 
 homogeneous term, one can determine a rational prefactor ${\cal R}$
for the master integral. Rescaling the master integral by this prefactor, 
one obtains a differential equation in which the coefficient of the 
homogeneous term is of ${\cal O}(\e)$. This equation can then be solved 
order by order in $\e$ by 
direct integration. The remaining constants of integration, which 
correspond to the boundary condition of the equation, are subsequently 
determined by using the fact that the master integral is regular 
in the whole kinematic plane with the exception of a few (at most three) 
branch cuts. 

For each master integral, we obtain a result of the form
\begin{equation}
\sum_i{\cal R}_i(y,z;\sabc,\e){\cal H}_i(y,z;\e)\; ,
\label{eq:ansatz}
\end{equation}
where the prefactor
${\cal R}_i(y,z;\sabc,\e)$ is a rational function of $y$ and $z$,
which is multiplied with an overall normalization factor to account for
the correct dimension in $\sabc$, while ${\cal H}_i(y,z;\e)$ is a Laurent
series in $\e$. The coefficients of its $\e$-expansion are then written
as the sum of 
two-dimensional harmonic polylogarithms up to a weight 
determined by the order of the series:
\begin{equation}
{\cal H}_i(y,z;\e) = \frac{\e^p}{\e^4} \sum_{n=0}^{4} \e^n
\left[ T_n(z) + 
\sum_{j=1}^n \sum_{\vec{m}_j\in V_{j}(z)} T_{n,\vec{m}_j}(z) H(\vec{m}_j;y)
\right] \; , 
\label{eq:calH}
\end{equation} 
where the $ H(\vec{m}_j;y) $ are two-dimensional harmonic polylogarithms
(2dHPL), which were introduced in~\cite{gr1} and 
$T_n(z)$, $T_{n,\vec{m}_j}(z) $ are $z$-dependent coefficients. 

\subsection{Harmonic polylogarithms}
Harmonic polylogarithms (HPL) were introduced in~\cite{hpl} as an
extension of the generalized polylogarithms of Nielsen~\cite{nielsen,bit}. 
They are 
constructed in such a way that they form a closed, linearly 
independent set under a certain class of integrations. We observe that the 
class of allowed integrations on this set can be extended {\it \a`{a} la
carte} by enlarging the definition of harmonic polylogarithms
in order to suit the needs of a particular calculation. We  made 
use of this feature by generalizing the one-dimensional HPL
of~\cite{hpl} to two-dimensional harmonic polylogarithms (2dHPL),
which appear in the solution of the differential equations for 
the three-scale master integrals discussed in\cite{gr1}. 
We briefly recall the 
HPL formalism~\cite{hpl}:
\begin{enumerate}
\item
The one-dimensional HPL $H(\vec{m}_w;x)$ is described by a $w$-dimensional 
vector $\vec{m}_w$ of parameters and by its argument $x$. $w$ is called 
the weight of $H$.
\item 
The HPL of parameters $(+1,0,-1)$ 
form a closed set under the class of integrations
\begin{equation}
\int_0^x \d x^{\prime} \left(\frac{1}{x^{\prime}}, 
\frac{1}{1-x^{\prime}},\frac{1}{1+x^{\prime}}\right) 
H(\vec{b};x^{\prime})\; .
\label{eq:intclass}
\end{equation}
\item
The HPL fulfil an algebra, such that 
a product of two HPL (with weights $w_1$ and $w_2$) 
of the same argument $x$ is a combination of HPL of argument 
$x$ with weight $w=w_1+w_2$.
\item The HPL fulfil integration-by-parts identities.
\item
The HPL are linearly independent.
\end{enumerate}
The generalization from one-dimensional to two-dimensional 
HPL starts from (\ref{eq:intclass}), which defines the class of 
integrations under which the HPL form a closed set. By inspection of 
the various 
 inhomogeneous terms of the $y$ differential equations for the 
three-scale master integrals discussed in this paper, we find that,
besides the denominators $1/y$ and $1/(1-y)$, also $1/(1-y-z)$ and 
$1/(y+z)$ appear. It is therefore appropriate to introduce an 
extension of the HPL, which forms a closed set under the class 
of integrations 
\begin{equation}
\int_0^y \d y^{\prime} \left(\frac{1}{y^{\prime}}, 
  \frac{1}{1-y^{\prime}},\frac{1}{1-y^{\prime}-z},\frac{1}{y^{\prime}+z}
\right) H(\vec{b};y^{\prime})\; .
\label{eq:twoddef}
\end{equation}
Allowing $(z,1-z)$ as components of the 
vector $\vec{m}_w$ of parameters does then define the 
extended set of HPL, which we call two-dimensional 
harmonic polylogarithms (2dHPL). They retain all properties 
of the HPL, in particular the algebra and the linear independence. 

Two-dimensional harmonic polylogarithms can be expressed in
terms of Nielsen's generalized polylogarithms up to weight 3, which 
is the maximum weight appearing in the divergent terms of two-loop 
four-point functions with one leg off-shell. These relations 
are tabulated in~\cite{gr1}.
At weight 4,
only some special cases relate to  generalized polylogarithms. 

\section{Summary of recent results}

For two-loop four-point functions with massless internal propagators
and  all legs on-shell, which are 
relevant for example in the next-to-next-to-leading order
calculation of two-jet production at hadron colliders, all master
integrals have been calculated over the past two
years. The calculations  were performed using the Mellin--Barnes 
method~\cite{onshell1} and the differential equation 
technique~\cite{onshell2}. The resulting master integrals can be 
expressed in terms of Nielsen's generalized polylogarithms. 
Very recently, these master integrals were already applied in the 
calculation of 
two-loop virtual corrections to Bhabha scattering~\cite{m1}
 in the limit of vanishing 
electron mass and to quark--quark scattering~\cite{m2}.

In~\cite{gr1}, we have used the differential equation approach 
to compute all master integrals for two-loop four-point functions 
with one off-shell leg. Earlier partial results on these functions were 
obtained in~\cite{glover,smirnew}, and a purely numerical approach to 
these functions was presented in~\cite{num}.
Our results~\cite{gr1} for these master
integrals are in terms of two-dimensional 
harmonic polylogarithms. All 2dHPL appearing in the divergent parts of 
the master integrals can be expressed in terms of 
Nielsen's generalized polylogarithms of suitable non-simple arguments, while 
the 2dHPL appearing in the finite parts are one-dimensional integrals over 
generalized polylogarithms. An efficient numerical implementation of these 
functions is currently being worked out. Our results
correspond to the kinematical situation of a $1\to 3$ decay, their analytic 
continuation into the region of $2\to 2$ scattering processes requires 
the analytic continuation of the 2dHPL, which is outlined in~\cite{gr1}.  

These four-point two-loop master integrals 
with one leg off-shell are a crucial ingredient to 
 the virtual next-to-next-to-leading order corrections to processes 
such as three-jet production in electron--positron annihilation, 
two-plus-one-jet production in deep inelastic scattering and 
vector-boson-plus-jet production at hadron colliders.

\section{Outlook}

Owing to numerous technical developments in the past two years,
virtual two-loop corrections to four-point amplitudes
are now becoming available for a variety of phenomenologically 
relevant processes. One must however keep in mind that these 
corrections form only one part of a full next-to-next-to-leading order 
calculation, which also has to include the one-loop corrections 
to processes with one soft or collinear real parton~\cite{onel}
 as well as 
tree-level processes with two soft or collinear partons~\cite{gg}. 
Only after summing all these contributions (and including 
terms from the renormalization of parton distributions 
for processes with partons in the initial state), the divergent terms cancel 
among one another. The remaining finite terms have to be combined 
into a numerical programme implementing the 
experimental definition of jet observables and event-shape variables. 
A first calculation involving the above features 
was presented for case of photon-plus-one-jet 
final states in electron--positron annihilation in~\cite{gg},
thus demonstrating the feasibility of this type of calculations.

\Acknowledgments

TG would like to acknowledge a travel grant from the Deutsche 
Forschungsgemeinschaft  (DFG) under contract number Ge1138/2-1.

\end{document}